\documentclass[twocolumn,english,aps,prl,showpacs]{revtex4}
\usepackage[T1]{fontenc}
\usepackage[latin9]{inputenc}
\usepackage{amsthm}
\usepackage{amsmath}
\usepackage{graphicx}
\usepackage{setspace}
\usepackage{amssymb}
\usepackage{esint}
\usepackage[english]{babel}

\begin{document}

\title{Expansion of 1D polarized superfluids: The FFLO state reveals itself}

\author{Hong Lu$^{1}$, L. O. Baksmaty$^{1}$, C. J. Bolech$^{2,1}$, and Han
Pu$^{1}$ }

\affiliation{$^{1}$Department of Physics and Astronomy, Rice University, 6100
Main street, MS-61, Houston, TX 77005.\\ 
$^{2}$Department of Physics, University of Cincinnati, 345 Clifton Court, ML-11, Cincinnati, OH 45221-0011.}
\date{\today}
\begin{abstract}
We study the expansion dynamics of a one dimensional polarized Fermi gas after sudden release from confinement using both the mean-field Bogoliubov-de Gennes and the numerically exact Time-Evolving Block Decimation methods. Our results show that experimentally observable spin density modulations directly related to the presence of a Fulde-Ferrel-Larkin-Ovchinnikov (FFLO) state develop during the expansion of the cloud, providing incontrovertible evidence of this long-sought state.
\end{abstract}
\maketitle

Since the introduction of the Bardeen-Cooper-Schrieffer (BCS) theory,
physicists have speculated on the fate of the superconducting pairing
correlation in the presence of a polarizing effect. This could arise
from a mass imbalance of the pairing fermions such as in color superconductivity
or in the vicinity of magnetic impurities within conventional superconductors.
The FFLO (Fulde-Ferrel-Larkin-Ovchinnikov) \cite{FF_original,LO_original,machida}
proposal suggests that in such circumstances the condensation will
occur from pairs with finite center-of-mass momenta. Despite decades
of work \cite{chandra,clogston,casalbuoni}, this state has not been
unambiguously observed. Although recent experiments \cite{yean_liao}
in one dimension (1D) confirmed important aspects of the phase diagram
\cite{Orso,paata}, conclusive evidence of the FFLO phase was not
obtained. We show here that during a non-equilibrium expansion, the
polarized 1D superfluid develops strong signatures in the density
profiles of the paring species which are a direct consequence of the
FFLO crystalline order and constitute incontrovertible evidence. 

We focus on a polarized degenerate Fermi gas confined to a 1D harmonic
trap. In general, according to the Mermin-Ho-Wagner theorem, a 1D
superfluid system cannot support superfluidity and would possess,
at best, algebraically decaying long range order at zero temperature
($T=0$). However for the finite systems that we study here, there
is copious theoretical evidence that FFLO correlations occur and are
fairly robust \cite{drummond_1d,Parish_Huse_mueller,fabian,casula_ceperly,ueda_tezuka,Guan,lee and guan,torma}.
We also note that the experiments use not a single 1D trap but a loosely
coupled array which allows tuning of the inter-tube coupling and thus
makes it possible to study the 3D to 1D crossover physics \cite{yean_liao}.
Although a partially polarzied phase was observed through direct imaging
in the experiment, it is quite clear from recent work that the FFLO
correlations \textit{do not} leave a detectable signature on the ground
state density profiles. Thus the character of the partially polarized
phase remains unknown. 

We consider a gas of $N$ fermionic particles each of mass $m$ with
two spin projections labeled by $\sigma=(\uparrow,\downarrow)$ confined
to a cigar-shaped harmonic trap. Consistent with experimental reality
\cite{yean_liao,Jin,mit_science,rice_science,mit_nature,Salomon},
we assume that the inter-particle interaction arises from a broad
feshbach resonance and is amenable to exquisite control. In these
systems, the ratio of the radial $\omega_{r}$ and axial $\omega_{z}$
trapping frequencies which define the anisotropy of the trap $\lambda=\omega_{r}/\omega_{z}$
can be made so large that the Fermi energy $E_{F}$ associated with
the axial dynamics of the trap $N\hbar\omega_{z}$ and the temperature
$k_{B}T$, are both much smaller than the energy level spacing of
the radial confinement $\hbar\omega_{r}$ i.e., $N\hbar\omega_{z},k_{B}T<<\hbar\omega_{r}$ \cite{yean_liao}.
Due to extremely rarerified nature of the gas, there are virtually
no spin relaxation processes and the particles interact via $s$-wave
scattering $g_{{\rm 1D}}\delta(z)$ . Furthermore, in addition to
the total number $N$, the total polarization of the cloud $P=(N_{\uparrow}-N_{\downarrow})/(N_{\uparrow}+N_{\downarrow})$
can also be varied through independent control of the number of particles
in each spin projection $N_{\sigma}$. Formally this system is described
by a Hamiltonian $\hat{H}=\int dz\,(H_{0}+H_{I})$ with : \begin{eqnarray}
H_{0}(z) & = & \sum_{\sigma}\psi_{\sigma}^{\dagger}\left[-\frac{\hbar^{2}}{2m}\frac{\partial^{2}}{\partial z^{2}}+V_{{\rm trap}}\left(z\right)-\mu_{\sigma}\right]\psi_{\sigma}\nonumber \\
H_{I}(z) & = & g_{1D}\psi_{\uparrow}^{\dagger}(z)\psi_{\downarrow}^{\dagger}(z)\psi_{\downarrow}(z)\psi_{\uparrow}(z)\label{eq:basic_hamiltonian}\end{eqnarray}
 where $\psi_{\sigma}(z)$ and $\mu_{\sigma}$ represent the fermionic
field operators and the chemical potential of atomic species with
spin $\sigma$ and $V_{{\rm trap}}(z)=\frac{m}{2}\omega_{z}^{2}z^{2}$.
We define the Fermi energy, radius, momentum and temperature as $E_{\textit{F}}=N$,
$z_{\textit{F}}=\sqrt{2E_{\textit{F}}}$, $k_{\textit{F}}=\sqrt{2E_{\textit{F}}}$
and $T_{F}=E_{F}$ and measure the relative strength of the interaction
with the ratio ($\gamma$) of the interaction ($\epsilon_{I}$) and
the kinetic ($\epsilon_{k}$) energy densities . In the limit of weak
interaction $\epsilon_{I}\sim g_{{\rm 1D}}\rho(z)$ and $\epsilon_{k}\sim\rho^{2}(z)$
yielding: \begin{equation}
\gamma=g_{{\rm 1D}}/\rho\label{eq:gamma}\end{equation}

\begin{figure*}
\includegraphics[width=0.75\textwidth]{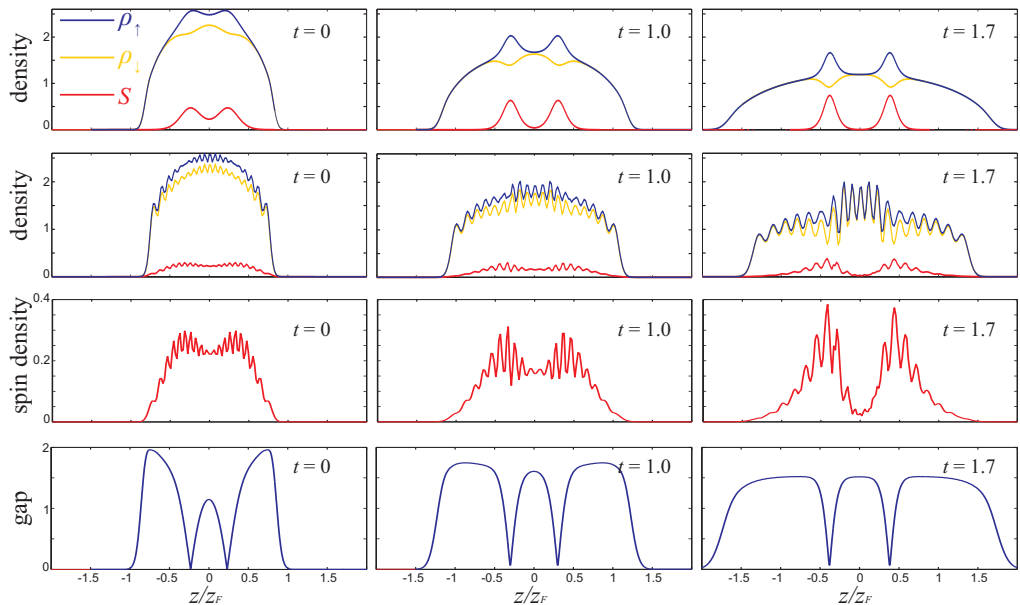}

\caption{\label{fig:p_050_g8_n40_exp}(color online)The expansion of sample with $N=40$,
$P=0.05$ and $g_{{\rm 1D}}=-8.0$. From left to right, each column
represents snapshots of the expansion dynamics at $t$=0.0, 1.0, 1.7
$(1/\omega_{z})$. Row 1 displays the density profiles. In each plot,
we show $\rho_{\uparrow}$, $\rho_{\downarrow}$ and $S=\rho_{\uparrow}-\rho_{\downarrow}$
obtained from BdG calculation. Row 2 is the same as Row 1 except that
the results are obtained from TEBD calculation. Row 3 shows the spin
densities $S(z)$ from the TEBD. Finally in Row 4 we plot the amplitude
of the superfluid gap $|\Delta|$ from the BdG calculation.}

\end{figure*}

Our calculations are done using two methods with distinct but complementary
advantages. First is the Time Evolving Block Decimation (TEBD) \cite{vidal}
(See Supplemental Material at for details of methods), an exact approach that retains all important correlations.
Second is the mean-field Bogoliubov-de Gennes (BdG) method, an effective
theory approach which retains only the two point correlations and
describes the spin densities $\rho_{\sigma}(z)$ and the superfluid
gap $\Delta(z)$ through quasi-particle wavefunctions. The BdG has
the advantage that, when correct, it provides a clear picture of the
dynamics of the pairing field $\Delta(z)=g_{{\rm 1D}}\langle\psi_{\uparrow}(z)\psi_{\downarrow}(z)\rangle$
in direct association with the particle densities $\rho_{\sigma}(z)=\langle\psi_{\sigma}^{\dagger}\left(z\right)\psi_{\sigma}\left(z\right)\rangle$.
However, although the BdG has been observed to give a very good description
of 1D samples at weak interaction \cite{drummond_1d}, we do not expect
this trend to extend from moderate to strong interaction. On the other
hand the TEBD method provides a stringent check for the observed phenomena
in the BdG approach. In both cases we work at $T=0$ and employ a
canonical approach which fixes $N$ and $P$.

To observe the FFLO state, experiments must verify crystalline order
in $\Delta(z)$ or alternatively, that the average center-of-mass
momentum of the pairs $\langle n_{k}\rangle$ is proportional to the
separation of the Fermi surfaces $\langle n_{k}\rangle\propto k_{\uparrow}-k_{\downarrow}$.
In 1D this relationship can be recast in terms of the spin density
$S(z)=\rho_{\uparrow}(z)-\rho_{\downarrow}(z)$ as $\langle n_{k}\rangle\propto\pi\int_{L}S(z)dz/L$,
where $L$ is the size of the partially polarized region. Recently
a number of authors \cite{Kun,fabian,casula_ceperly} have suggested the
measurement of the pair momentum distribution function $n_{k}$ as
the most promising avenue to detecting the finite center-of-mass momentum
$q$ of the pairs. These suggestions are extrapolations from equilibrium
studies where $n_{k}$ shows peaks at $k=\pm q$ in contrast to the
peak at $k=0$ expected for regular BCS pairing. However, we are not
aware of analyses of $n_{k}$ accounting for the interacting nature
of the expansion dynamics and in particular how well this signal will
be preserved. This is particularly important for 1D given that $\gamma$
increases during expansion {[}see Eq.~(\ref{eq:gamma}){]}. In this
study we explore the possibility of finding a signal directly in real
space. Our calculations reveal that: 
\begin{itemize}
\item Upon axial expansion, strong accents develop in the spin density profiles. 
\item The position of these accents exactly coincide with the nodes in the
pair correlation function and represent {\em prima facie} evidence
of FFLO correlations. 
\item The strength of this signal increases with $\gamma$ and decreases
with polarization, being strongest when the spin excitations are gapped. 
\item The accents in the spin density move much more slowly than the edge
of the cloud. 
\end{itemize}

In Fig.~\ref{fig:p_050_g8_n40_exp} dramatic accents in the spin
densities are observed as the expansion of the cloud proceeds. Through
a comparison of the density plots with the corresponding gap parameter
$|\Delta(z)|$ (bottom row in Fig.~\ref{fig:p_050_g8_n40_exp}) one
can make a key observation: \textit{The position and growth of the
spin density accents respectively coincide with the nodes and amplification
of $|\Delta(z)|$}. Furthermore, these spin density accents (or the
order parameter nodes) move much slower during the expansion compared
to the edge of the whole cloud.

To understand this phenomenon, it is helpful to first layout some
broad features of the ground state utilizing the phase diagram for
a homogeneous system together with the local density approximation
(LDA) \cite{Orso,paata,ueda_tezuka}. Under LDA, the trapped system
can be regarded as locally homogeneous with chemical potential defined
by: $\mu(z)=\mu_{\sigma}-V_{{\rm trap}}(z)$. There are two regimes
to be considered \cite{yean_liao,drummond_1d,Parish_Huse_mueller,fabian,ueda_tezuka}
depending on whether $P$ is smaller or larger than a critical polarization
$P_{c}$. For $P<P_{c}$, we obtain an FFLO state at the center of
the trap surrounded by fully paired BCS wings at the edges. Here the
BdG calculation tells us that there is exactly one excess spin bound
to each of the nodes of the order parameter and the FFLO state is
analogous to a band insulator of the \textit{relative motion} between
the unpaired and paired particles. The ground state represented in
Fig.~\ref{fig:p_050_g8_n40_exp} is within this regime and density
accents represent the localization of unpaired spins at the nodes
of $\Delta$. During the time of flight, the excess spins are kept
pinned to the nodes of the order parameter and become more tightly
bound. The dramatic effects observed occur when this localization
couples with the average enhancement of $|\Delta|$ implied by an
increasing $\gamma$ as the density drops during expansion {[}see
Eq.~(\ref{eq:gamma}){]}; a uniquely 1D phenomenon. Henceforth we
refer to these accents as node signatures.

\begin{figure}
\includegraphics[width=0.3\textwidth]{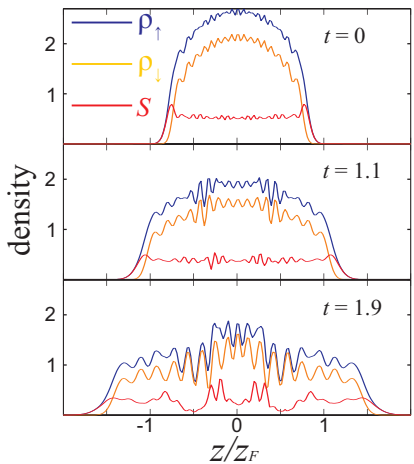}

\caption{\label{fig:p_150_g8_n40_exp}Density profiles, obtained from TEBD
calculation, during the expansion of a sample with $N=40$, $P=0.15$
and $g_{{\rm 1D}}=-8.0$.}

\end{figure}

For $P>P_{c}$, the FFLO state still remains at the center in the
ground state, but the wings exclusively contain the majority spin
component. In this regime, there are more excess spins than nodes
of $\Delta$, and consequently they are less tightly bound. Here we
expect the node signatures to be less dramatic which is confirmed
in Fig.~\ref{fig:p_150_g8_n40_exp}. In particular, the spin accents
near the edges are not well resolved. We can therefore conclude that
the best place to observe the node signature is at $P<P_{c}$ where
the signal is enhanced by both a large separation of the nodes and
greater contrast with the background density. We note that the value
of $P_{c}$ increases with $|g_{{\rm 1D}}|$ implying a sizable observation
window at strong interactions where experiments are conducted.

\begin{figure}
\includegraphics[width=0.4\textwidth]{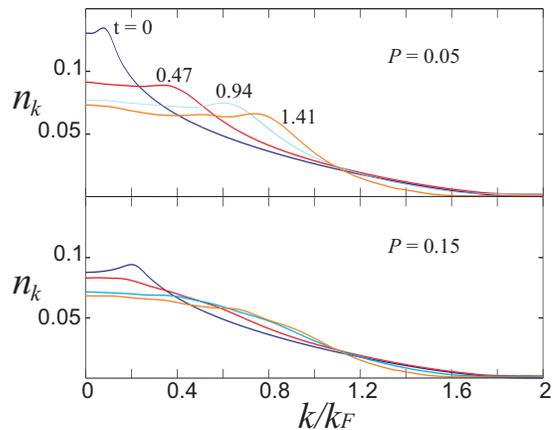}

\caption{\label{corr}Pair momentum distribution at two different polarization
for $g_{{\rm 1D}}=-8$ and $N=40$. In each panel, we display $n_{k}$
for different times. Counting from the left, the curves correspond
to $t=0$, 0.47, 0.94 and 1.41, from top to bottom. In both cases
the momentum peaks representing the FFLO state disappear from the
plot during the expansion. }

\end{figure}

At equilibrium the FFLO correlation appear as peaks in the pair momentum
distribution $n_{k}$ defined by: \begin{equation}
n_{k}=\frac{1}{L}\,\int\int dzdz{'}\, e^{ik(z-z{'})}\, O(z,z{'})\,,\label{pair}\end{equation}
 where $O(z,z{'})\equiv\langle\psi_{\uparrow}^{\dagger}(z)\psi_{\downarrow}^{\dagger}(z)\psi_{\downarrow}(z')\psi_{\uparrow}(z')\rangle$
is the two-point correlation function. In Fig.~\ref{corr} we observe
the effects of interaction on this signature during the expansion.
At sufficiently long time, $n_{k}$ no longer possesses peaks at finite
momentum.

\begin{figure}
\includegraphics[width=0.3\textwidth]{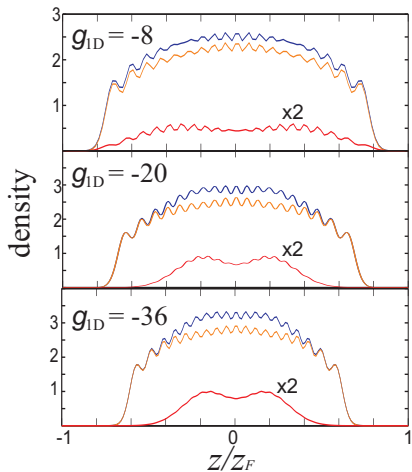}

\caption{\label{GS}Ground state density profiles in trap, with $N=40$, $P=0.05$
and for different interaction strengths $g_{{\rm 1D}}$. In each plot,
we show $\rho_{\uparrow}$, $\rho_{\downarrow}$ and $S$ obtained
from TEBD calculation. For clarity, the spin density $S$ is magnified
by a factor of 2.}

\end{figure}

One may wonder whether the node signatures can be observed in {\em in situ} density profiles of a trapped
cloud with sufficiently large interaction strength. To answer this, we show
in Fig.~\ref{GS} the density profiles of a trapped system for $g_{\rm 1D}=-8$, $-20$ and $-36$. (Note that for the experiment reported in Ref.~\cite{yean_liao}, $g_{\rm 1D} \sim -50$ for the central tube.) One can see that the modulation depth of the spin density of a trapped cloud is
not very sensitive to $g_{{\rm 1D}}$. This is in sharp contrast to the BdG
calculation where the spin density modulation is indeed enhanced as
$\gamma$ is increased --- an indication of the invalidity of the mean-field theory for strong interaction. In the exact calculation, the localization
of excess spin at large $|g_{{\rm 1D}}|$ is counter-balanced by increased
quantum fluctuations neglected in the mean-field theory. Therefore,
the dramatic emergence of node signatures is a unique feature of expansion
dynamics.

\begin{figure}
\includegraphics[width=0.45\textwidth]{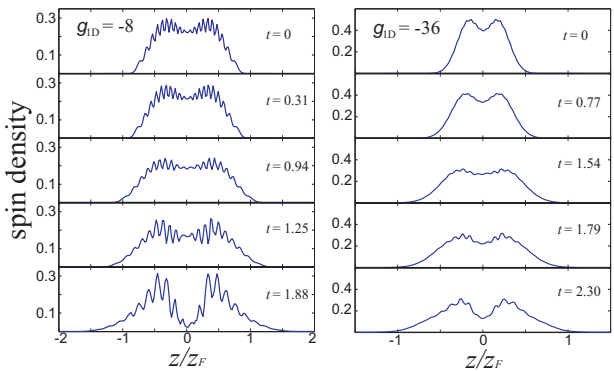}

\caption{\label{Spin} Expansion profiles for two different samples with $N=40$,
$P=0.05$ but at different interaction strengths $g_{{\rm 1D}}$.
In each plot, we plot the TEBD result for $S$. In both cases, the
modulation depth of the spin density first reduces and then strengthens
during expansion.}

\end{figure}

Finally, we address the question of the effect of the interaction strength
in Fig.~\ref{Spin} where the spin densities in an expanding cloud are shown for two sets of interaction strength. Though the results from the strong and weak interaction are qualitatively similar, the spin accents start to develop earlier for the case of smaller $g_{{\rm 1D}}$. This could play an important role in practice when finite lifetime of the system must be taken into account.    

In conclusion, we have investigated the expansion dynamics of polarized Fermi superfluid in 1D using both the BdG and TEBD methods. Our results predict that strong spin density modulations which can be readily observed in experiment, emerge during expansion and provide direct evidence of the FFLO state. Apart from the pair momentum distribution function described above, other methods \cite{inter} have been proposed in the literature to detect FFLO. However they all rely on interferometric techniques requiring two fermionic superfluids, one of them being the FFLO state. Our proposal, in contrast, only requires the FFLO cloud itself and hence is significantly simpler. In a more general context, our work shows that quantum dynamics of low-dimensional atomic gases is highly non-trivial and deserve a more thorough study in the future.         
 
Part of the numerical calculations for this work was performed at NERSC, Navy DSRC, ARL, AFRL and
the ARSC. We thank Eric Mueller, Randy Hulet, Micheal Wall, Yean-an Liao and S.
Bhongale for several illuminating discussions. This work is supported
by the ARO Award W911NF-07-1-0464 with the funds from the DARPA OLE
Programm, the Welch foundation (C-1669, C-1681) and the NSF.

\clearpage

\section{Methods - Supplementary material}

This system of $N=N_{\uparrow}+N_{\downarrow}$ is described by a
Hamiltonian $\hat{H}=\int dz \,(H_{0}+H_{I})$ with non-interacting
$(H_{0})$ and interaction $(H_{I})$ energy densities given by: \begin{eqnarray}
H_{0}(z) & = & \sum_{\sigma}\psi_{\sigma}^{\dagger}\left[-\frac{\hbar^{2}}{2m}\frac{\partial^{2}}{\partial z^{2}}+V_{trap}\left(z\right)-\mu_{\sigma}\right] \psi_{\sigma}\,,\nonumber \\
H_{I}(z) & = & g_{1D}\psi(z)\psi_{\downarrow}^{\dagger}(z)\psi_{\downarrow}(z)\psi_{\uparrow}(z)\,,\label{eq:basic_hamiltonian}\end{eqnarray}
 where $\psi_{\sigma}(z)$ represent the Fermionic field operators,
$m$ the mass and $\mu_{\sigma}$ the chemical potential of atomic
species with spin $\sigma$. The 1D effective coupling constant $g_{1D}<0$
is expressed through a relationship with the 3D scattering length
$a_{3D}$ by \cite{olshanii}: $g_{1D}=\frac{2\hbar^{2}a_{3D}}{ma_{l}(1-Aa_{3Dd}/a_{l})}$.
Here $a_{l}$ is the oscillator length and $A=\xi(1/2)/\sqrt{2}$.
We work in 'trap' units: $m=\omega_{z}=\hbar=k_{B}=1$.

\subsection*{BdG Calculation}

We treat $\hat{H}$ within the mean-field Bogoliubov-de Gennes (BdG)
approach for which there are many excellent references \cite{drummond_3d}.
Here we simply state the BdG equations for the pair wave functions
$u_{j}(z)$ and $v_{j}(z)$ which decouple $\hat{H}$ :

\begin{equation}
\left[\begin{array}{cc}
H_{\uparrow}^{S}-\mu_{\uparrow} & \Delta({z})\\
\Delta({z}) & -H_{\downarrow}^{S}+\mu_{\downarrow}\end{array}\right]\left[\begin{array}{c}
u_{j}\\
v_{j}\end{array}\right]=E_{j}\left[\begin{array}{c}
u_{j}\\
v_{j}\end{array}\right],\label{mean_field_ham}\end{equation}

where $E_{j}$ is the associated energy. Despite this, the BdG treatment
has been shown to yield qualitatively reliable answers \cite{drummond_3d}.
In accordance with Fermionic commutation relations, the quasi-particle
amplitudes are normalized as: $\int dz \,|u_{j}(z)|^{2}+|v_{j}(z)|^{2}=1.$
In terms of which the Gap $\Delta(z)$ and the free energy $\Omega$,
may be written as : \begin{eqnarray}
\Delta(z) & = & U\,\sum_{j=1}^{\infty}u_{j}(z)v_{j}^{*}(z)f\:(E_{j})\,,
\label{eq:gap_uv}
\end{eqnarray}
where$f(E)$ represents the Fermi-Dirac distribution function:
$f(E)=1/(e^{E/k_{B}T}+1)$. We follow a convention that $N_{\uparrow}>N_{\downarrow}$,
we define $k_{\textit{F}}^{\uparrow\downarrow}=\sqrt{2\mu_{\uparrow\downarrow}}$
and the FFLO wave number by $q_{\,0}=k_{F}^{\uparrow}-k_{F}^{\downarrow}$.

Our theoretical framework is encapsulated within Eqs.~(\ref{mean_field_ham})
and (\ref{eq:gap_uv}) and we discretize the system of Eq.~(\ref{mean_field_ham})
using a piece-wise linear finite element basis which ensures the continuity
of both $u(z)$ and $v(z)$. A reduction
of Eq.~(\ref{mean_field_ham}) into even and odd parity states about
$z=0$ is possible due to anticipated reflection symmetry
of $\Delta$ about this axis. Nevertheless, each independent sub-block
with distinct parity presents a very large eigenvalue
problem because of the slow convergence of Eq.~(\ref{eq:gap_uv}). The slow convergence
is tackled using a hybrid BdG-semi-classical strategy similar to Ref.~\cite{drummond_3d}.
Starting from an initial state Eq.~(\ref{mean_field_ham}) is iteratively
solved to self-consistency using the modified Broyden's method \cite{Johnson}.
We work in a canonical formalism which keeps $N$ and the total polarization
$P=(N_{\uparrow}-N_{\downarrow})/N$ fixed through the number equations
$N_{\sigma}=\int dz \,\rho_{\sigma}(z).$

\subsection*{TEBD calculations}

To implement TEBD formalism, we employ a 1D Fermi-Hubbard Hamiltonian
to approximate the continuum quasi-1D polarized Fermi gases in harmonic
traps: \begin{equation}
\begin{split}
H=&-J\,\underset{{\scriptscriptstyle \sigma}}{\sum}\overset{{\scriptscriptstyle L}}{\underset{{\scriptscriptstyle i=2}}{\sum}}(c_{i,\sigma}^{\dagger}c_{{\scriptscriptstyle i-1,\sigma}}+h.c.)+U\overset{{\scriptscriptstyle L}}{\underset{{\scriptscriptstyle i=1}}{\sum}}n_{i,\uparrow}n_{i,\downarrow}\\
&+\overset{{\scriptscriptstyle L}}{\underset{{\scriptscriptstyle i=1}}{\sum}}V_{i}(n_{{\scriptscriptstyle i,\uparrow}}+n_{{\scriptscriptstyle i,\downarrow}})\,,\label{hubbard}
\end{split}
\end{equation}
where $L$ is the number of discretized lattice sites, $c_{{\scriptstyle {\scriptscriptstyle i.\sigma}}}^{\dagger}$,
$c_{{\scriptstyle {\scriptscriptstyle i.\sigma}}}$ are respectively
the creation and annihilation operators for spin $\sigma$ particles
at $i$th lattice site, $J$ is the hopping amplitude between the
neighboring sites, and $U$ is the onsite interaction strength between
two unlike spins. The connection between the Fermi-Hubbard Hamiltonan
(\ref{hubbard}) and the Hamiltonian (\ref{eq:basic_hamiltonian})
upon which the BdG calculation is based can be seen as follows: In the trap units
we mentioned above, the hopping amplitude $J=\frac{L^{2}}{2l^{2}}$,
where $l$ is the total length of system (in our dimensionless units).
From these, the parameters $\frac{U}{J}=2g_{{\rm 1D}}l/L$ and $\frac{V_{i}}{J}=(\frac{l}{L})^{4}(i-\frac{L}{2})^{2}$
are choosen accordingly. In our calculation, typically we choose $L=300\thicksim400$.
With these properly choosen characteristic parameters, the discretized
Hubbard Hamitonian can be trusted to represent a continuum system
as have been previously shown \cite{ueda_tezuka-1}.

The TEBD algorithm utilizes the Schmidt decomposition and the convergence
of the simulation is mainly controlled by the so-called Schmidt rank
$\chi$, which is the number of eigenvalues retained when truncating
the Hibert space. In our calculation, since the computation time scales
as $\chi^{3}$, the optimal value of $\chi\sim{100-150}$ is chosen to ensure
the convergence is good enough when comparing with the results with
higher $\chi$. Another souce of error comes from the Trotter-Suzuki
expansion for the time evolution operator. To reduce it, we adopt
fifth order Trotter-Suzuki expansion in our calculation while choosing
a small enough time step based on self-consistent stability test.

\end{document}